\documentclass[pra,twocolumn,showpacs]{revtex4}
\usepackage{graphicx}
\graphicspath{{fig/}{}}

\usepackage{bm}

\newcounter{Fig}

\newcommand{\be}{\begin{equation}}
\newcommand{\ee}{\end{equation}}

\begin{document}

\title{Stable higher-charge discrete vortices in hexagonal optical lattices}

\author{Kody J.H. Law, P.~G. Kevrekidis}

\affiliation{Department of Mathematics and Statistics, University of Massachusetts,
Amherst MA 01003-4515}

\author{Tristram J. Alexander, Wieslaw Kr{\'o}likowski, and Yuri S. Kivshar}

\affiliation{Nonlinear Physics Center and Laser Physics Center, Research School of Physical Sciences and
Engineering, Australian National University, Canberra ACT 0200, Australia}


\begin{abstract}
We show that double-charge discrete optical vortices may be completely stable
in hexagonal photonic lattices where single-charge vortices always exhibit dynamical
instabilities.  Even when unstable the double-charge vortices typically have a much 
weaker instability than the single-charge vortices, and thus their breakup occurs 
at longer propagation distances.
\end{abstract}


\maketitle

\section{Introduction}

Some of the most spectacular experiments in the field of nonlinear light propagation
in periodic photonic potentials relate to the properties of vortices and vortex
flows~\cite{PIO}. Self-trapped phase singularities of optical fields have been observed
experimentally in the form of single-charge discrete optical vortices in square photonic lattices~\cite{Neshev,moti,Bartal:2005-53904:PRL}.  In addition, many of the theoretical and experimental
studies demonstrated that higher-charge discrete vortices are unstable, similar to the well-studied
homogeneous nonlinear systems~\cite{PIO}.

In this work, we study single- and double-charge discrete optical vortices in
non-square periodic photonic lattices~\cite{honey,kouk,rosberg2,gaid,tja2,bernd}. In particular,
in the framework of the continuous nonlinear model of optically-induced lattices generated
in saturable nonlinear media, we analyze the existence, stability and dynamical properties
of discrete optical vortices for the case of {\em hexagonal optical lattices}. We obtain the
somewhat counter-intuitive result that double-charge discrete vortices in such lattices
appear to be far more robust and structurally stable than single-charge vortices.  We verify 
this finding by demonstrating numerically the generation of a double-charge vortex with 
realistic experimental parameters.

It is particularly important to highlight that although
our results will be given with a view towards applications
in photorefractive crystals, they are not only relevant to that setting 
but also directly applicable to two-dimensional hexagonal waveguide
arrays (e.g., in glass), showcased in recent experiments (see e.g.,
\cite{szameit} and references therein). Furthermore, they are likely
to have direct implications to other areas of physics, such as
Bose-Einstein condensates (BECs) in triangular lattices, the first experiments
of which have just been realized \cite{klaus}, or even Debye crystals
in dusty plasmas \cite{yannis}. 
Another key aspect of the generality of
our results is that they should {\it also} apply to honeycomb lattices.
Hence, the findings presented herein have a bearing on two of the
most fundamental non-square lattice two-dimensional configurations.

\begin{figure}[h]
\noindent\includegraphics[width=\columnwidth]{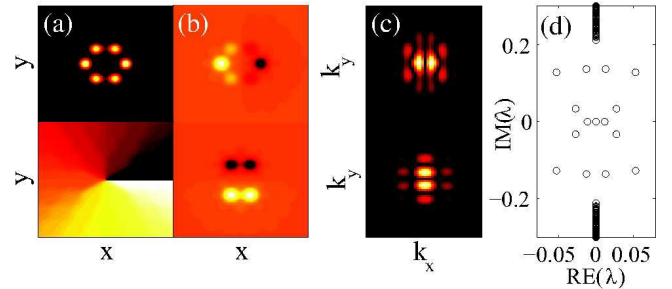} \caption{
(Color online) Example of an always unstable single-charge discrete optical vortex for $\beta=-0.76$ (marked by a circle in Fig.~\ref{fig2}).  (a) Intensity (top) and
phase (bottom); (b) real (top) and imaginary (bottom) components;
(c) absolute value of the corresponding Fourier transforms; (d) spectrum of
the linearized equation displaying the linear instability of the configuration due to the presence of positive real parts in the eigenvalues $\lambda$ in the spectrum.
}
\label{fig1}
\end{figure}

\begin{figure}[h]
\noindent\includegraphics[width=\columnwidth]{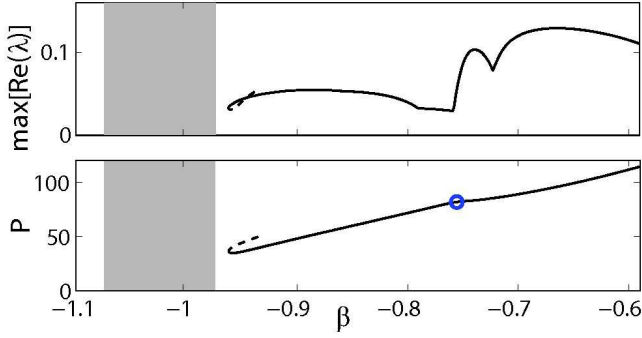} \caption{(Color online) Family of single-charge vortices vs. propagation constant $\beta$. Top: maximum real part of the linear stability spectrum. Bottom: power $P=\int_\infty^\infty U^2dxdy$.  The circle corresponds to the discrete vortex given in 
Fig.~\ref{fig1}.  The dashed line indicates another unstable branch which, for larger $\beta$ bifurcates into different configurations.
}
\label{fig2}
\end{figure}

\begin{figure}[h]
\noindent\includegraphics[width=\columnwidth]{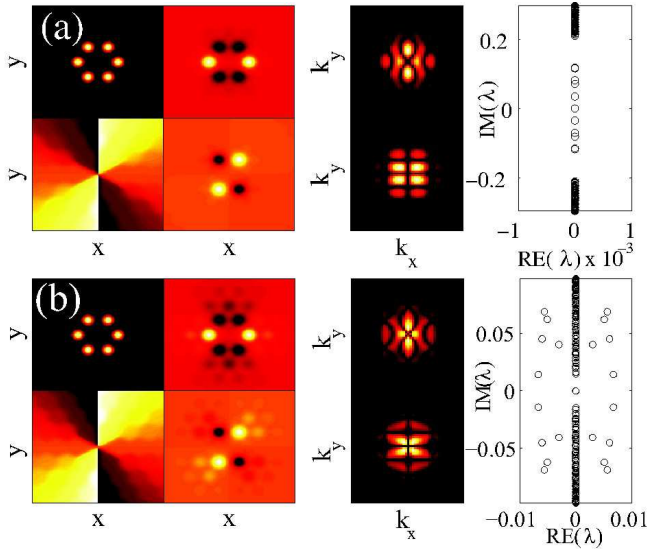} \caption{
(Color online) Examples of (a) stable and (b) unstable double-charge discrete optical vortices for $\beta=-0.76$ and $\beta=-0.96$ respectively (marked respectively by the circle and square in Fig.~\ref{fig4}).  The layout of the panels is the same as in Fig.~\ref{fig1}.
}
\label{fig3}
\end{figure}

\begin{figure}[h]
\noindent\includegraphics[width=\columnwidth]{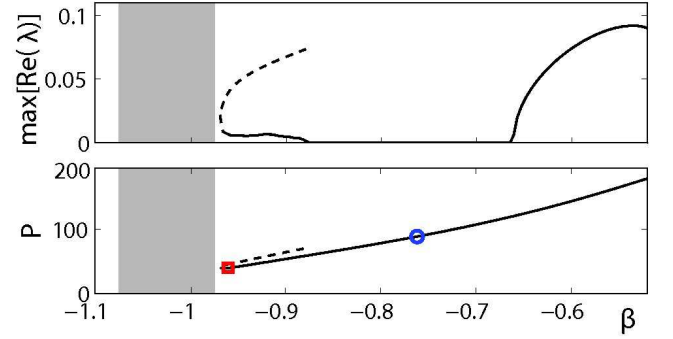} \caption{(Color online) Family of double-charge vortices vs. propagation constant $\beta$. Top: maximum real part of linear stability spectrum 
[when nonzero, this denotes instability]. Bottom: power $P=\int_\infty^\infty U^2dxdy$.  The circle and square correspond to the stable and unstable discrete vortex configurations shown in Fig.~\ref{fig3} ((a) and (b) respectively).  The dashed line indicates an unstable
branch which, for larger $\beta$ bifurcates into different configurations.
}
\label{fig4}
\end{figure}

\section{Theoretical Setup}

We study beam propagation through a self-focusing nonlinear medium
in the presence of a two-dimensional hexagonal lattice by employing the continuum model
with a saturable nonlinearity. To render our setting completely
amenable to the experimentally accessible regime, we use the
theoretical model of a photorefractive nonlinear medium, which is known
to exhibit strong saturable nonlinearity~\cite{rosberg2}.
Polarization anisotropy of the nonlinear photorefractive response
enables one to  optically imprint various types of refractive index
modulation (optical lattice) which can then be probed by an external
beam~\cite{efrem}. Then the  propagation of this beam  in the
presence of an optically-induced hexagonal refractive index pattern is
governed by the following normalized evolution equation
\begin{eqnarray}
i \frac{\partial u}{\partial z} + D \Delta_{\perp} u -
\frac{\gamma \, u}{1+ I_p(x,y) + |u|^2}=0, \label{hex_eq6}
\end{eqnarray}
where $u(x,y; z)$ is the normalized amplitude of the electric field,
$z$ is the propagation coordinate, $\Delta_{\perp}$
denotes the transverse Laplacian with respect to $(x,y)$, $D$ is the
relevant diffraction coefficient, and $\gamma$ is the material
parameter which is positive or negative depending on whether the
nonlinearity is of focusing or defocusing character. 
The function
$I_p(x,y)=I_g |\exp(i k x)+\exp(-ikx/2-iky\sqrt{3}/2) +\exp(-ikx/2 +
i k y\sqrt{3}/2)|^2$ represents  the three-wave interference pattern
that induces the hexagonal lattice.  The lattice and beam intensities are normalised in units of the dark irradiance of the crystal, $I_b$.  Throughout this work we use the following experimentally realistic values for the system parameters: $D = z_s\lambda/(4\pi n_0 x_s^2) = 18.015$ (for laser wavelength in vacuum $\lambda = 532$nm, and average refractive index of the medium $n_0 = 2.35$), $\gamma = 2.36$, $I_g = 0.49$, $k = 4\pi/3d$ with a lattice period $d = 30\mu$m, and where the dimensions $(x,y;z)$ are in units of $x_s = y_s = 1\mu$m and $z_s = 1$mm respectively (see Ref.~[7] for further details).

\begin{figure}[h]
\noindent\includegraphics[width=\columnwidth]{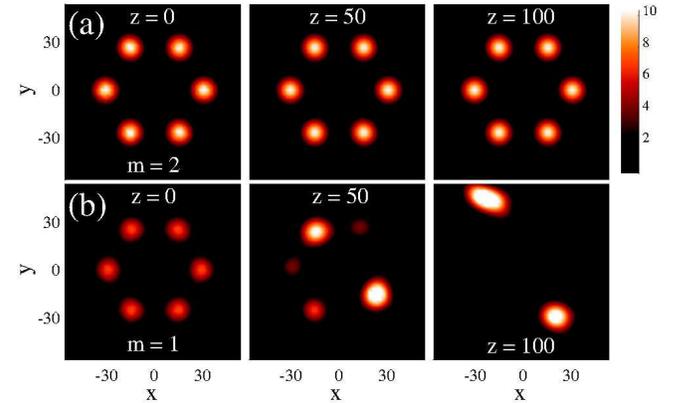} \caption{(Color online) The top three panels (a) depict the evolution of a stable double-charge vortex
configuration after a random perturbation with amplitude 5\%
of the initial amplitude.  The bottom three panels (b) are the evolution
of a single-charge vortex configuration.  In both cases $\beta=-0.7$.  The color bar on the right provides a scale of the intensity (note the intensities of the single-charge vortex are lower relative to (a) initially and saturated on this scale after break-up).
}
\label{fig5}
\end{figure}

\begin{figure}[h]
\noindent\includegraphics[width=\columnwidth]{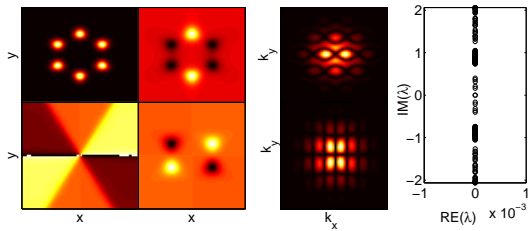} 
\caption{(Color online) The same set of panels as in Fig. \ref{fig1} 
except for a stable double-charge vortex in a honeycomb lattice.}
\label{fig6}
\end{figure}


\begin{figure}[h]
\noindent\includegraphics[width=\columnwidth]{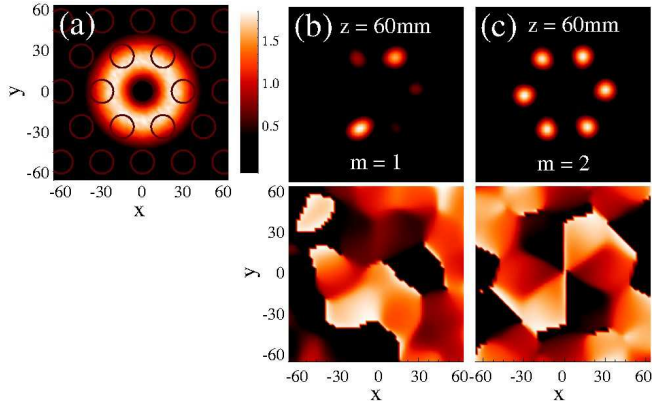} \caption{(Color online) (a) Input beam intensity profile relative to the lattice (position of lattice intensity maxima are shown as rings).  The intensity is given by the color-bar on the immediate right.  Appearance of the beam at $z = 60$mm with different initial vortex phases (intensity not to scale of color-bar); (b) single-charge vortex, (c) double-charge vortex.  Top panels: intensities; bottom panels: phase.
}
\label{fig7}
\end{figure}

We look for stationary solutions in the form $u(x,y; z) = U(x,y) \exp (i\beta z) \exp (im\phi)$, where
$U$ is real, $\beta$ is the propagation constant, $\phi$ is the vortex phase, and $m$ is the vortex charge. We solve the resulting nonlinear equation numerically, and the major results are summarized in what follows.

\section{Numerical Results}

We begin by considering the simplest six-site vortex structure, that of a single-charge ($m=1$) discrete vortex.  Figure \ref{fig1} illustrates a typical example, while Fig.~\ref{fig2} shows the single-charge vortex linear stability (top) and power (bottom) as a function of the propagation constant in the semi-infinite band gap of the periodic potential.  A positive real part of an eigenvalue in the linear stability spectrum leads to exponential growth of the corresponding linear excitation mode, and therefore to instability of the vortex.  Somewhat surprisingly we can see in Fig.~\ref{fig2}(top) that the single-charge vortex has an eigenvalue with a positive real part across its entire region of existence, and therefore the single-charge vortex is {\em always unstable}.

In contrast we find that double-charge vortices may be {\em stable} (see Fig.~\ref{fig3}(a)), and even where unstable the instability is weaker than the single-charge case (see Fig.~\ref{fig4}).  In fact, as we can see in Fig.~\ref{fig4}(top) the double-charge vortex has a wide parametric interval 
where it is completely stable (from $-0.92<\beta<-0.65$), while outside this range it is
unstable due to weak oscillatory instabilities (complex unstable eigenvalues, as evidenced by the spectrum in Fig.~\ref{fig3}(b)).  We note that neither the single- nor double-charge discrete vortex families degenerate into
a linear Bloch mode, as one can observe from the saddle-node bifurcation that
occurs close to the edge of the first band of the linear spectrum in both Fig.~\ref{fig2} and Fig.~\ref{fig4}.  The various unstable single- and double-charge vortices which occur along the upper dashed branch in each figure are not discussed here.  The typical evolution of the stable and unstable vortices is illustrated in Figs.~\ref{fig5}(a,b).  Even though the single-charge vortex is lower in power than the double-charge vortex, break-up of the former into single-site fundamental discrete solitons occurs around $z=50$, while the double-charge vortex has been propagated to $z=1000$ with no sign of instability.

To further our theoretical understanding, 
we employ a discrete model. In the latter
 the analytically tractable anti-continuum limit can be used, for which discrete vortex 
solutions can be explicitly constructed and a detailed stability analysis 
can be performed, as has been done for square lattices~\cite{peli2d}.
In such a setting we consider the six site configuration with topological
charge $m$ over the contour, which takes the form 
$u_{j}=\exp(i n\phi_{j}) \exp(i\beta z)$, 
where $\phi_j=2 \pi j m/6$ and $j=1,\dots,6$ for the
six sites constituting the relevant contour. It is straightforward
to see that this configuration yields non-trivial phase profiles for
$m=1$ and $m=2$. For these structures, according to the framework
of \cite{peli2d}, the fundamental vortex will be {\it unstable} due to two
double real eigenvalue pairs and a single real eigenvalue pair whereas
the $m=2$ configuration may be {\it stable}.
These general results may also be physically 
understood as a consequence of the 1D modulational instability (MI) 
results~\cite{mi} along the 1D (with periodic boundary conditions) six-site 
contour of the vortex. Such MI considerations
 predict that configurations where adjacent 
sites have less than a $\pi/2$ phase difference (i.e. a single-charge vortex) 
will be unstable, while those with more than a 
$\pi/2$ phase difference (the double-charge case) will be stable.  

It is important to point out here that, as the above discrete 1D contour
analysis suggests, our results 
can qualitatively be extended to other cases where there exists a 
six-site closed contour, as e.g. in
the so-called {\it honeycomb} lattice in which each index maximum has three 
neighboring maxima instead of six.  A typical example of a stable 
double-charge  
vortex in a honeycomb lattice is presented in Fig. \ref{fig6}.  
Furthermore, extending our consideration of the 1D six-site 
contour to the case of a defocusing nonlinearity, one can apply a
so-called {\it staggering} transformation  {\it along the contour}, 
$U_j = u_j (-1)^j$.  Substitution of this expression in the discrete 
equation transforms the model from defocusing to focusing (and vice versa).  This amounts
to translating the phase of every other node along the contour by $\pi$ 
and, hence, transforming an $m=1 (m=2)$ focusing vortex to an $m=2 (m=1)$ defocusing vortex respectively, suggesting a corresponding stability exchange.

\section{Experimental proposal}

Finally, we consider the generation of double-charge vortices and suggest parameters for their experimental observation.  For our particular lattice parameters we find that generation of stable double-charge vortices is possible over a wide range of input beam intensities and profiles, at least within our isotropic medium approximation.  We consider a Laguerre-Gaussian input beam with the profile shown in Fig.~\ref{fig7}, kept as constant as possible as the input phase is changed, with maximum intensity $\sim 1.8I_b$.  In the subsequent evolution we see break up of the beam into single-site discrete solitons if the initial phase corresponds to a single-charge vortex (Fig.~\ref{fig7}(b)), while with an initial double-charge vortex phase we see stable generation of the discrete double-charge vortex (Fig.~\ref{fig7}(c)).  Output at $z=60$mm is shown, however we have seen no sign of instability in the generated double-charge vortex at a distance of $z=500$mm.

Based on the above considerations, we believe that inputs
of the type associated with $m=2$ should be sustained during
propagation not only by hexagonal crystals in photorefractive
media, but also by two-dimensional hexagonal waveguide
arrays (e.g., in glass), showcased in recent experiments~\cite{szameit}.
Importantly also, similar results are theoretically expected
and have been numerically confirmed (data not shown here) to be
valid in the case of honeycomb lattices in such media.


\section{Conclusions \& Outlook}

We have studied the existence, stability, dynamics and generation of single- and double-charge
discrete optical vortices in two-dimensional hexagonal optical lattices in the framework of a continuum
nonlinear model for photorefractive nonlinearity.
We have found that, in contrast to square lattices, double-charge vortices can be stable, 
while single-charge vortices are always unstable. Our main finding constitutes a general result for
both hexagonal and honeycomb lattices 
that we expect to be verified experimentally.

There are numerous directions along which it would be interesting to
continue the present study. For example, it would be relevant to extend
our analysis to the defocusing case for
which our discrete theory predicts that the results should be
inverted (i.e., that the $m=1$ should be potentially stable, while
the $m=2$ will be unstable). Another direction of interest would
be to attempt to generalize such studies to genuinely three-dimensional,
non-square lattice settings (e.g., in fcc, bcc or hcp crystals)
and observe how realistic three-dimensional excitations may behave
in these classes of models. Some of these directions are under 
present consideration and will be reported in future publications.

\section*{Acknowledgements}

P.G.K. thanks the Nonlinear Physics Center and the Laser Physics Center 
at the Australian
National University for the warm hospitality during his visit to Canberra. 
P.G.K. also gratefully acknowledges support from NSF-DMS-0349023,
NSF-DMS-0505663, NSF-DMS-0619492, and NSF-DMS-0806762,
as well as from the Alexander von
Humboldt Foundation.  This work has been
supported by the Australian Research Council through the Discovery 
Project scheme.


\begin{thebibliography}{99}

\bibitem{PIO} A.~S. Desyatnikov, Yu.~S. Kivshar, and L. Torner, Prog. Optics {\bf 47},
Ed. E. Wolf (Elsevier, Amsterdam, 2005), pp. 291--391, and references therein.

\bibitem{Neshev} D.~N. Neshev, T.~J. Alexander, E.~A. Ostrovskaya, Yu.~S. Kivshar, H.
Martin, I. Makasyuk, and Z. Chen, Phys. Rev. Lett. {\bf 92}, 123903 (2004).

\bibitem{moti} J.~W. Fleischer, G. Bartal, O. Cohen, O. Manela, M. Segev, J. Hudock, and
D.~N. Christodoulides, Phys. Rev. Lett. {\bf 92}, 123904 (2004).

\bibitem{Bartal:2005-53904:PRL} G. Bartal, O. Manela, O. Cohen, J. W. Fleischer, and M. Segev, Phys.
Rev. Lett. {\bf 95}, 053904 (2005).


\bibitem{gaid} P.G. Kevrekidis, B.A. Malomed, and Yu.B. Gaididei,
Phys. Rev. E {\bf 66}, 016609 (2002).

\bibitem{kouk} V. Koukouloyannis and R.S. MacKay, J. Phys. A: Math. Gen. {\bf 38}, 1021 (2005).

\bibitem{honey} O. Peleg, G. Bartal. B. Freedman, O. Manela,
M. Segev and D.N. Christodoulides, Phys. Rev. Lett. {\bf
98}, 103901 (2007).

\bibitem{rosberg2} C.R. Rosberg, D.N. Neshev, A.A. Sukhorukov,
W. Krolikowski and Yu.S. Kivshar, Opt. Lett. {\bf
32}, 397 (2007).

\bibitem{tja2} T.J. Alexander, A.S. Desyatnikov, and Yu.S. Kivshar, Opt. Lett.
{\bf 32}, 1293 (2007).

\bibitem{bernd} B. Terhalle, T. Richter, A.S. Desyatnikov, D.N. Neshev, W. Krolikowski,
F. Kaiser, C. Denz, and Yu.S. Kivshar, Phys. Rev. Lett. {\bf 101}, 013903 (2008).

\bibitem{szameit} A. Szameit, Y. V. Kartashov, F. Dreisow, M. Heinrich,
V.A. Vysloukh, T. Pertsch, S. Nolte, A. Tunnermann, F. Lederer, L. Torner,
Opt. Lett. {\bf 33}, 633 (2008).

\bibitem{klaus} See e.g., C. Becker, P. Soltan-Panahi,
J. Kronjager, S. Stellmer, K. Bongs and K. Sengstock,
``Spinor BEC in Triangular Optical Lattices'', IB1\_1 IQEC (2007).

\bibitem{yannis} V. Koukouloyannis and I. Kourakis,
Phys. Rev. E {\bf 76}, 016402 (2007).

\bibitem{efrem} N. K. Efremidis, S. Sears, D. N. Christodoulides, J. W. Fleischer,
and M. Segev, Phys. Rev. E {\bf 66}, 46602 (2002).

\bibitem{peli2d} D.E. Pelinovsky, P.G. Kevrekidis, and D.J.
Frantzeskakis, Physica D {\bf 212}, 20 (2005).

\bibitem{mi} Yu.S.~Kivshar and M.~Peyrard, Phys.~Rev.~A {\bf 46}, 3198 (1992); D.N. Christodoulides and R.I. Joseph, Opt. Lett. {\bf 13}, 794 (1988).


\end{thebibliography}
\end{document}